\documentclass[graybox]{svmult}

\usepackage{mathptmx}       % selects Times Roman as basic font
\usepackage{helvet}         % selects Helvetica as sans-serif font
\usepackage{courier}        % selects Courier as typewriter font
\usepackage{type1cm}        % activate if the above 3 fonts are
\usepackage{makeidx}         % allows index generation
\usepackage{graphicx}        % standard LaTeX graphics tool
\usepackage{multicol}        % used for the two-column index
\usepackage[bottom]{footmisc}% places footnotes at page bottom

\usepackage{amsmath}
\usepackage{graphicx}
\usepackage{amsfonts}
\makeindex             % used for the subject index
\makeindex
\begin{document}

\title*{Supertree Construction: Opportunities and Challenges}
\titlerunning{Supertree Estimation}
% Use \titlerunning{Short Author list} for an abbreviated version of
% your list of authors,  if the original one is too long
\author{Tandy Warnow}
\institute{Tandy Warnow \at University of Illinois at Urbana-Champaign \email{warnow@illinois.edu}
%\and Some Co-Author \at Some Institution \email{someaddress@illinois.edu}
}
\maketitle

\abstract{
Supertree construction is the process by which a set of phylogenetic trees, each on a subset of the overall set $S$ of species, is combined into a tree on the full set $S$.
%The main reason that supertrees are constructed is that phylogeny estimation is a computationally hard problem, as nearly all methods with good accuracy are based on NP-hard optimization criteria and do not scale well to large datasets; hence, the main use of supertree estimation is  to enable the construction of very large trees.
The traditional use of supertree methods is the assembly of a large species tree from previously computed smaller species trees; however, supertree methods are also used to address large-scale tree estimation using divide-and-conquer (i.e., a dataset is divided into overlapping subsets, trees are constructed on the subsets, and then combined using the supertree method). 
Because most supertree methods are  heuristics for NP-hard optimization problems, the use of supertree estimation on large datasets is challenging, both  in terms of scalability  and   accuracy.
In this chapter, we describe the current state of the art in supertree construction and the use of supertree methods in divide-and-conquer strategies.
Finally, we identify 
directions where future research could lead to improved supertree methods.
%The abstract should be at most 200 words, and probably should not include any references to other papers.
}

\section{Introduction}
\label{sec:1}
%I plan to write a survey paper about supertree methods or else about species tree estimation methods. Both these topics are relevant to 
% Bernard Moret's work.

The estimation of phylogenetic trees, whether  gene trees (that is, the phylogeny relating the sequences found at a specific locus in different species) or species trees,  is a precursor to many
downstream analyses, including protein structure and function prediction,  the detection of positive selection, co-evolution, human population genetics, etc.~(see \cite{hillis_application_1994} for just some of these applications). 
Indeed, as Dobzhanksy said, ``Nothing in biology makes sense except in the light of evolution" \cite{Dobzhansky}.

Standard  approaches for estimating phylogenies from a given set of molecular sequences
typically involve two steps: first, a multiple sequence alignment  is computed, and then a tree is computed on the multiple sequence alignment. 
Both steps are computationally intensive on large datasets, but for somewhat different reasons.
Multiple sequence alignment is approached using many different techniques, and generally runs in polynomial time, but when the input set of sequences 
is both large and highly heterogeneous (in the sense that many insertions and deletions have occurred in the history relating the sequences), then the alignments can become
extremely large -- much larger than the input set -- presenting challenges if the available memory is insufficient for the alignment itself.
Highly accurate large-scale multiple sequence alignment can be difficult to obtain with standard methods, but new approaches (largely based on divide-and-conquer) have
been developed that enable multiple sequence alignment methods to scale with high accuracy to ultra-large datasets \cite{sate2009,upp,Nute2016}.

Tree estimation presents a different challenge: while some polynomial time methods have been developed for tree estimation (e.g., neighbor joining and its variants \cite{NJ,bionj,ninja}, FastME \cite{fastme-2015}, and other 
distance-based methods), simulation studies suggest that  computationally intensive methods, such as heuristics for maximum likelihood (ML) and Bayesian MCMC,  produce more accurate
phylogenies \cite{wang-protein-msa}.
Maximum likelihood codes are more frequently used than Bayesian MCMC methods, for the simple reason that they can be used on large datasets.
Advances in techniques for maximum likelihood (discussed in Chapters XX and YY in this volume) have led to codes with substantially improved scalability, including RAxML \cite{stamatakis_raxml-ni-hpc_2006}, PhyML \cite{phyml}, IQTree \cite{IQtree}, and FastTree-2 \cite{fasttree-2}.
Despite these advances, the estimation of single gene phylogenies 
using the best ML  methods is typically limited to at most a few thousand sequences (and much fewer for multi-gene phylogenies), and even these can be very computationally intensive.
For example, our own analyses of  single-gene datasets with 10,000 or more sequences can take weeks for RAxML, one of the leading maximum likelihood codes, to converge to good local optima.
Thus, truly large-scale maximum likelihood phylogeny estimation is still not feasible in practice. 

%Thus,  large-scale tree estimation is challenging for multiple reasons, including the challenges in estimating highly accurate multiple sequence alignments, and computing trees on these large alignments using either maximum likelihood or Bayesian methods.
Another challenge in large-scale tree estimation is ``heterotachy" \cite{Lopez-heterotachy-2002}, which is where  the statistical models that best characterize the evolutionary processes operating on the sequences (such as the rates of change between different nucleotides) vary across the tree, thus creating opportunities for model misspecification when standard statistical models are used in maximum likelihood or Bayesian phylogeny estimation.

Supertree methods, which combine trees on subsets of the taxon set into a tree on the full set of species, can be used to address these challenges \cite{Clann}.
The early supertree methods  largely focused on the calculation of supertrees for the idealized case where the source trees  could be combined into a species tree that agreed with them perfectly.
Since phylogeny estimation nearly always has some error, this requirement on the source trees is highly restrictive. 
Today,  there are many supertree methods that can handle conflict between the source trees and the construction of supertrees on biological datasets is a fairly common occurrence; for example, 
\cite{bininda-emonds_building_1999,jones_phylogenetic_2002,pisani_genus-level_2002,davies_darwins_2004,salamin_using_2004,grotkopp2004evolution,fernandez2005complete,pisani_supertrees_2007,jonsson_phylogenetic_2006,baker2009complete}  present species trees computed using supertree methods.
Supertrees also provide insight into a surprisingly large number of biological questions, as surveyed in \cite{bininda-emonds_supertree_2002}.

One of the earliest supertree methods to be developed is Matrix Representation with Parsimony, also referred to as MRP \cite{baum_combining_1992,ragan_phylogenetic_1992}.
MRP is so widely used that it has become synonymous with ``supertree method''.
Today, the development of new species tree methods is  an active research area in computational phylogenetics, and draws on discrete mathematics, computer science, and probability theory, and many new supertree methods have been developed, some of which are even more accurate than MRP.  

Yet more research is needed, as even the best supertree methods fail  to be highly accurate on large datasets with many thousands (and even tens of thousands) of species, which is where supertree methods are most needed \cite{bininda-emonds_supertree_2002}, and most cannot even run on datasets of these sizes.

%add citations
The purpose of this chapter is to explore the  challenges to scalability for existing supertree methods. 
We identify the major algorithmic approaches in supertree construction and the challenges of scaling those approaches to large datasets, and identify open problems where advances by computer scientists would potentially lead to practical supertree methods with good accuracy on large datasets.
We also discuss the use of supertree methods in divide-and-conquer strategies that aim to estimate phylogenetic trees on ultra-large datasets.
The basic theoretical material  for supertree construction is provided here in a brief form, and further details  can be found in \cite{warnow-book}.
\index{maximum likelihood tree estimation}
%Need cites to back that up.
\index{distance-based tree estimation}
 \index{supertree methods} 
 \index{Tree of Life}
 
\section{Background}

\subsection{Terminology}
We introduce the basic terminology used in supertree estimation.
\begin{definition}
The input to a supertree problem is a set $\mathcal{T}$ of trees.
The set $\mathcal{T}$ is referred to as a {\bf profile} and the individual trees in $\mathcal{T}$ are  referred to as {\bf source trees}. 
We let $\mathcal{L}(t)$ denote the leafset of tree $t$.
Then a supertree for $\mathcal{T}$ is a tree $T$ with one leaf for every   $s \in S = \cup_{t \in \mathcal{T}} \mathcal{L}(T)$
\end{definition}

The focus in this chapter is on unrooted source trees, since techniques for rooting phylogenies depend on careful selection of outgroup taxa and can introduce error into the supertree profile. 
Furthermore, although in some cases the source trees will have branch lengths or branch support values,  in the classical setting (and most common use case) the source trees come without any branch lengths or branch support values and are just ``tree topologies". 
Hence, the description of supertree methods in this chapter is for this simplest setting where the input set of source trees are unrooted binary trees without branch lengths or support values.

Finally, we assume that the source trees will differ from the true species tree (when restricted to the same taxon set) only due to {\em estimation error}.
Thus, we do not address the case where the source trees are gene trees (either estimated or true gene trees), as these can differ from the true species tree and from each other due to processes such as gene duplication and loss, incomplete lineage sorting, and horizontal gene transfer \cite{maddison1997}.
When the input profile consists of gene trees, then methods that explicitly address gene tree discord (such as 
phylogenomic ``summary methods", which combine estimated gene trees to produce an estimated species tree) are more suitable, and are discussed in Chapter \ref{sec:phylogenomics}, below. 
%Of course, supertree methods can be used as summary methods and vice-versa, but the evaluation protocols are different!

For a given profile $\mathcal{T}$, the best possible outcome is where there is a supertree $T$ that agrees with all the source trees.
We formalize this as follows:
\begin{definition}
We let $T|X$ denote the subtree of $T$ induced by the leaves in $X$.
If $T$ is a supertree for $\mathcal{T}$ and $t \in \mathcal{T}$ is a binary tree, we 
say that $T$ agrees with  $t$
if $T|\mathcal{L}(t)$ is isomorphic to $t$.
We also say that $T$ is a {\bf compatibility tree} for $\mathcal{T}$ if $T$ agrees with $t$
for all $t \in \mathcal{T}$.
When a compatibility tree exists for a set $\mathcal{T}$, we say that  the set $\mathcal{T}$ is  {\bf compatible}.
\end{definition}
However, usually source trees are not compatible (because tree estimation nearly always has some error), so that the objective is to find a  supertree that minimizes some total distance (or maximizes some total similarity score) to the input source trees \cite{thorleyview2003,Clann}.

\subsection{Representations of Trees}
Unrooted supertrees and input source trees can be represented in several different ways, and many of the supertree methods that have been developed are based on these representations.
Here we describe three basic techniques for representing trees: sets of bipartitions, sets of quartet trees, and additive distance matrices. 

\begin{definition}
Every edge in a tree $t$ defines a bipartition on the leafset of $t$ (i.e., when you delete the edge $e$ from $t$, you separate the leaves into two sets, $A_e$ and $B_e$, thus defining the bipartition $A_e|B_e$).  
We let {\bf Bip(t)} denote the set of bipartitions of the leafset of $t$ defined by the edges of $t$.
\end{definition}
It is not hard to see that the set $Bip(t)$ defines the tree $t$, and that $t$ can be reconstructed from $Bip(t)$ in polynomial time \cite{warnow-book}.

\begin{definition}
Every set $q$ of four leaves in an unrooted binary tree $t$ induces a quartet tree $t_q$, and hence $t$ defines the set {\bf Q(T)} of all its induced quartet trees.
\end{definition}
As with bipartitions, the set $Q(t)$ defines the tree $t$, and $t$ can be reconstructed from $Q(t)$ in polynomial time \cite{warnow-book}.
The last representation we present is based on distance matrices:

\begin{definition}
A matrix $D$ is said to be {\bf additive} if there is a tree $T$ with $n$ leaves and non-negative lengths on the edges so that $D_{ij}$ is the total of the edge lengths on the path  in $T$ between leaves $i$ and $j$.
\end{definition}
Given an additive matrix, there is a unique minimal edge-weighted tree corresponding to the additive matrix \cite{buneman1971}, and it can be constructed from the additive matrix in polynomial time (see \cite{waterman77} for the first such method, and see \cite{warnow-book} for others).

For a given unrooted tree $t$, there are several ways of representing $t$ using an  additive matrix.
For example, some source trees are computed using methods such as maximum likelihood or neighbor joining \cite{NJ}, which explicitly provide positive branch lengths on the edges of the tree; given such a tree, the matrix of leaf-to-leaf distances (summing all the branch lengths on the path) is by definition additive.
Other source trees come without branch lengths, and so using unit lengths on all the edges produces an additive matrix. 
%Another way of defining the distance between two leaves is called the ``internode distance", which is the number of internal nodes on the path between the pair of leaves. Each of these ways of defining a distance matrix creates an ``additive matrix" \cite{waterman76}, and 
Therefore, whether the source trees are either unweighted or have strictly positive branch lengths, they can be used to define additive matrices, and  are uniquely defined by their additive matrices.

Thus, every unrooted binary tree $t$ can be defined using either its set $Bip(t)$ of bipartitions, its set $Q(t)$ of quartet trees, or by an additive matrix $A$, and given any of these representations the tree $t$ can be reconstructed in polynomial time.
Each of these representations suggests approaches for constructing supertrees from a set $\mathcal{T}$ of source trees. 
For each of the following general categories of problems, the input is a set $\mathcal{T}$ of source trees and the objective is a supertree $T$ that is close to the source trees, with the measurement of distance based on one of these representations of source trees and the supertree.

\subsection{Bipartition-based supertree methods}

Three of the most well-known  optimization problems for supertree construction --
Matrix Representation with Parsimony, Robinson-Foulds Supertrees, and Maximum Likelihood Supertrees -- are based on bipartition encodings.
Furthermore, Matrix Representation with Likelihood is a new approach that also uses the bipartition encoding and has been shown to be competitive with  MRP.
Therefore, we will begin with this category of supertree methods.

\subsubsection{Matrix Representation with Parsimony}
%Matrix Representation with Parsimony (MRP) is
%Another such approach is to treat the bipartition encodings as discrete characters (with missing data for the species that do not appear in a given source tree), and then compute the maximum parsimony score of these characters on the supertree.
%This results in the 
We begin with
the well-known NP-hard supertree problem ``Matrix Representation with Parsimony" (MRP)
\cite{ragan_phylogenetic_1992,baum_combining_1992,bininda-emonds_mrp_2003,mrp}.
There are many variants on this problem, described in \cite{baker2009complete}, but the basic approach is the one that is most frequently used.
The input set $\mathcal{T}$ of  (source) trees is
represented  by a matrix (called the MRP matrix) with $0$s, $1$s, and $?$s, as follows.
\begin{definition}
Given a profile $\mathcal{T}$ of source trees, each edge $e$ in each tree $t \in \mathcal{T}$ defines a bipartition $ A_e|B_e$ on $\mathcal{L}(t)$.
The bipartition $A_e|B_e$ can be represented as a $n$-tuple (where $|S|=n$) where the $i^{th}$ entry has $0$ for the elements $s_i \in A_e$, $1$ for the elements $s_i \in B_e$, and $?$ for the elements in $S \setminus (A_e \cup B_e)$.
These $n$-tuples representing the bipartitions (taken across all the edges from all the source trees) form the columns in the MRP matrix.
Thus, the {\bf MRP matrix} has
one row for every species in $S = \cup_t \mathcal{L}(t)$ and one column for every internal edge in any tree in $\mathcal{T}$.
\end{definition}
Once the MRP matrix is computed, the rows of the matrix are treated as strings (all having the same length) that will represent the leaves of the tree we will seek, and  a  maximum parsimony tree is sought for the matrix:
%\subsubsection{The Maximum Parsimony Problem}
\begin{definition}
The {\bf cost} of a tree $T$ in which all the nodes are labelled by strings of length $k$ is the total of the Hamming distances on the edges of the tree, where
the Hamming distance between two $k$-tuples $s=(s_1, s_2, \ldots, s_k)$ and $s'=(s'_1, s'_2, \ldots, s'_k)$ is the number of positions $i$ for which $s_i \neq s'_i$.
The tree $T$ with the lowest achievable cost (among all ways of assigning sequences to its internal nodes) is said to be a {\bf maximum parsimony tree}.
\end{definition}
%Figure \ref{fig:mrp-clann} gives an example of the MRP matrix computed on four source trees, and the maximum parsimony tree found for that matrix.
%Tandy - uncomment this later

%The input to the Maximum Parsimony Problem is a matrix $M$ with one row for every species in $S$ and (without loss of generality) integer entries; hence, each species in $S$ is represented by a $k$-tuple of integers.
%The parsimony score of a fixed tree $T$ is computed by first assigning sequences of length $k$ to the internal nodes of $T$, and then adding up the Hamming distances on the edges. 
%Given a tree $T$ with the elements of $S$ at the leaves, it is possible to compute $k$-tuples at the internal nodes so that the total of the Hamming distances on the edges of the tree is minimized \cite{fitch_toward_1971,sankoff75} in polynomial time, using dynamic programming \cite{fitch_toward_1971,sankoff75}. 
%Hence, it is possible to compute the best possible parsimony score of a given rooted tree topology (with leaves labeled by the sequences in $S$) in polynomial time.

%The Maximum Parsimony problem seeks the tree that has the best possible parsimony score.
%Since scoring a fixed tree is  solvable in polynomial time, one approach to solving maximum parsimony is to explicitly score each tree and then return the tree with the best parsimony score. 
%However, this approach uses exponential time as there are an exponential number of trees on $n$ leaves (specifically there are $(2n-3)!!$ unrooted binary trees on $n$ leaves). 
Exact solutions to maximum parsimony are unlikely to be achievable in polynomial time, since
finding the best possible tree is NP-hard \cite{foulds_steiner_1982}.
For this reason, heuristic searches (using standard maximum parsimony software, such as PAUP* \cite{swofford_paup*_2002} or TNT \cite{tnt2008}) are used to find good solutions (i.e., local optima) to maximum parsimony.
The running time of these heuristics is mainly impacted by $n$, the number of sequences in the input, as the number of trees grows exponentially in $n$, but the number of sites (sequence length) also impacts the running time as the time to calculate the score of a given tree is linear in the number of sites. 
%There are several software packages that are available for maximum parsimony, with TNT \cite{tnt03,tnt2008} and PAUP* \cite{swofford_paup*_2002} two of the most popular.
Note that when the source trees are compatible, then the optimal solutions to MRP are compatibility trees, and hence an exact solution MRP will return a compatibility tree in that case;
this is one of the positive aspects of the MRP approach to supertree estimation.
In addition, simulation studies have generally found MRP to be among the most accurate supertree methods, with MRP referred to as   the ``gold standard'' among supertree methods  developed as of 2011 \cite{brinkmeyer2011polynomial}.

%\begin{figure}[b]
%\center
%\includegraphics[scale=.5]{supercat-mrp.png}
%\caption{(From \cite{tourasse2007supercat}) We show a pipeline that can be used to construct a supertree from a set of genes. }
%\label{fig:supercat}       % Give a unique label
%\end{figure}

%Tandy - uncomment this
%\begin{figure}[b]
%\center
%\includegraphics[scale=.4]{mrp-clann.eps}
%\caption{(From \cite{Clann}) The MRP pipeline: construct source trees (here, gene trees on different loci), then construct the MRP matrix from the set of source trees, and then  run maximum parsimony heuristics to obtain the MRP supertree.  For this particular example, note that the MRP tree induces each of the four source trees, so that they are all compatible. Note also that the characters (i.e., columns) in the MRP matrix are compatible on the MRP tree. }
%\label{fig:mrp-clann}       % Give a unique label
%\end{figure}

%Figure \ref{fig:supercat} shows a common pipeline that can be used when the source trees are different gene trees (as opposed to species trees estimated on subsets of the taxa).
%Note that in this pipeline,  the MRP supertree is a consensus tree of all the best trees found using the maximum parsimony heuristics on the MRP matrix, and that the branches of the consensus tree are then annotated with statistical support values.

%\subsubsection{Other uses of the MRP matrix}
The MRP approach is just one of the approaches for constructing supertrees from the MRP matrix, and other optimization problems can be considered for the same matrix representation of the input source trees.
For example, Matrix Representation with Likelihood (MRL). \cite{mrl}, Matrix Representation with Compatibility (MRC) \cite{rodrigo1993comment}, and Matrix Representation with Flipping \cite{chen_flipping_2003} have also been proposed.
These are also NP-hard problems, and so heuristics are used to find good solutions. 

Of these methods, only MRL has been shown to provide accuracy comparable to (and sometimes exceeding) MRP \cite{mrl}; hence, MRL deserves some additional discussion.

\subsubsection{Matrix Representation with Likelihood}
The MRL optimization problem takes the same input matrix as MRP, and randomizes the $0$s and $1$s to avoid bias. 
Then,  a solution to maximum likelihood under the two-state symmetric sequence evolution model is sought using  RAxML \cite{stamatakis_raxml-ni-hpc_2006}. 
As shown in \cite{mrl},  supertrees computed using MRL  were more accurate than supertrees computed using MRP, whether TNT or PAUP* heuristics were used, 
and MRL scores showed somewhat stronger correlations with tree accuracy than MRP scores.  
Furthermore, RAxML was reasonably fast on these datasets, finishing in less time than the parsimony heuristics that were used.

%\subsection{Benefits and Disadvantages of MRP and its variants}

\subsubsection{Robinson-Foulds Supertrees}
We continue with the Robinson-Foulds Supertree \cite{bansal_rf_2010, Chaudhary-RFsupertree}.

\begin{definition}
The {\bf Robinson-Foulds (RF) distance} between two binary trees $T_1$ and $T_2$ on the same leafset is $|Bip(T_1) \triangle Bip(T_2)| = |Bip(T_1) \setminus Bip(T_2) | + |Bip(T_2) \setminus Bip(T_1)|$ \cite{robinson_comparison_1981}.
When $T_1$ and $T_2$ have different leafsets, then we define the RF distance between $T_1$ and $T_2$ to be $RF(T_1|L,T_2|L)$, where $L = \mathcal{L}(T_1) \cap \mathcal{L}(T_2)$ \cite{CottonWilkinson2007}.
Given profile $\mathcal{T}$ of source trees, the RF distance between a supertree $T$ and $\mathcal{T}$ is 
$\sum_{t \in \mathcal{T}} RF(T,t)$. 
A {\bf Robinson-Foulds Supertree} for $\mathcal{T}$ is a supertree $T$ that has the minimum RF distance to $\mathcal{T}$.
\end{definition}
The Robinson-Foulds  Supertree problem is NP-hard, since determining if the profile $\mathcal{T}$ is compatible is NP-hard. 
Algorithms for the Robinson-Foulds Supertree problem include MulRF \cite{mulrf}, PluMiST \cite{plumist-journal}, and FastRFS  \cite{fastrfs}.

\subsubsection{Likelihood-based supertree methods}

Supertree estimation can be framed as a statistical inference problem that treats the estimated supertree $T$ as the model species tree that generates the set $\mathcal{T}$ of observed source trees.
In this setting, the objective is a maximum likelihood supertree, i.e., a tree $T$ such that $Pr(\mathcal{T}|T)$ is maximized. %that is most likely to generate the observed profile of source trees.

The initial formulation of this approach is due to Steel and Rodrigo \cite{SteelRodrigo}, who defined the probability of a source tree, given a model species tree, as a function of the Robinson-Foulds distance between the two trees, a development that was met with great excitement in the literature \cite{CottonWilkinson2009}. 
Two methods that have been developed that are based on the Steel and Rodrigo model are
\cite{Akanni-MLsupertree}, which attempts to compute the maximum likelihood supertree, and the Bayesian method in
 \cite{Akanni-Bayesian}.
Furthermore, as noted in \cite{BryantSteel-distribution}, there is a (potentially close) relationship between optimal solutions to the Robinson-Foulds Supertree  and Maximum Likelihood Supertree problems, suggesting that good solutions to one problem might be pretty good solutions to the other problem.
Hence, methods for the Robinson-Foulds Supertree may provide a good approximation to the maximum likelihood supertree problem under the Steel and Rodrigo model.
%Other approaches have also been proposed \cite{Ronquist,Posada-BayesianSupertree, Guenomu2017}.

\subsection{Quartet-based supertree methods}

Since each unrooted tree $t$ can be defined by its set $Q(t)$ of quartet trees,  quartet-based supertree optimization problems have been posed. 
Here we describe a very simple such optimization problem.

\begin{definition}
The {\bf Maximum Quartet Support Supertree} problem is defined as follows. 
The input is a profile $\mathcal{T}$ of source trees, and we seek the supertree $T$ that maximizes $qsim(T,\mathcal{T})$, where  $qsim(T,\mathcal{T}) = \sum_{t \in \mathcal{T}} |Q(T) \cap Q(t)|$.
\end{definition} 

The Maximum Quartet Support Supertree problem is NP-hard, since  even the decision problem of determining whether a set of unrooted source trees is compatible is NP-hard \cite{steel-compatibility}.
Furthermore, it remains NP-hard even for the special case where every source tree is on the full set $S$ of taxa \cite{LafondScornavacca2016}.
Note that quartet-based supertree problems can also be formulated as minimizing the quartet distance to the source trees, as these are equivalent.

One approach to solving this problem is to represent every source tree by its set of quartet trees, use those sets of quartet trees to produce a single quartet tree for every  four leaves (e.g., by majority vote), and then construct a tree on the full set of species from the constructed set of quartet trees\footnote{This approach has been used to construct species trees from gene trees under the multi-species coalescent; an example of such a method is  the population tree in BUCKy \cite{larget-bioinf2010}, but see also \cite{warnow-book} for others.}.

The last step in this process, where quartet trees are combined into a single tree on the entire dataset, is called
 ``quartet amalgamation".
The most well-known quartet amalgamation methods are   Quartet Puzzling \cite{strimmer_quartet_1996} and  Quartets MaxCut \cite{snir_quartets_2008}, but other methods have also been developed with good theoretical and empirical performance \cite{quartet-joining-conf,reaz-quartet,piaggio-talice_quartet_2004,ben1998constructing}. 
Some of these quartet  amalgamation methods are based on weighted versions of the standard quartet amalgamation problem,  where the input set of quartet trees have weights on them, perhaps reflecting the level of confidence in the quartet tree.
Weighted quartet amalgamation methods include Weight Optimization \cite{ranwez_quartet-based_2001} and Weighted Quartets MaxCut \cite{avni2014weighted}, and these weighted versions tend to provide better accuracy than unweighted quartet amalgamation methods. 
Since quartet compatibility is itself NP-complete, optimization problems for quartet amalgamation are NP-hard; however, many approximation algorithms theoretical results about quartet amalgamation have been developed
\cite{berry2000practical,BerryGascuel2000,bocker2000algorithmic,jiang_polynomial-time_2001,gramm2003fixed,AlonSnirYuster2014,LafondScornavacca2016}.
 %Add quartet cleaning methods, possibly?

 %, and ASTRAL \cite{astral,astral-2}.

Quartet-based supertree methods have been developed and studied, and often shown to have very good accuracy,  sometimes matching or improving on heuristics for MRP, generally considered the best supertree method for various criteria \cite{piaggio-talice_quartet_2004,swenson-amb-2011,AvniYonaCohenSnir2018}.
However, the default usage of these methods typically depends on using all the quartet trees, and hence explicitly will run in $\Omega(n^4)$ time, where there are $n$ leaves; hence, quartet-based supertree methods will typically be inapplicable to large datasets.
Some quartet amalgamation methods  can be run on just a subset of the quartet trees, which enables quartet-based supertree methods to be applied to sparse samples of the quartet trees. 
A study \cite{swenson-amb-2011} evaluating the impact of sampling quartets on supertrees computed using Quartets MaxCut \cite{snir_quartets_2008}  showed that some sampling strategies produced supertrees that were competitive with MRP, but the most accurate supertrees were obtained when all quartet trees are used  rather than a subset of the quartet trees.

\subsection{Distance-based supertree methods}
%cite SDM!!!
Distance-based supertree optimization problems are also popular \cite{Willson,criscuolo_sdm_2006,lapointe1997average,Willson,brinkmeyer2011polynomial}. 
As with quartet-based methods, one approach is to seek an additive matrix that is as close as possible to the set of additive matrices defining the source trees.
To make this concrete, for each source tree $t$ an additive distance matrix $D^t$ is computed. 
Then, given an additive distance matrix $D^T$ (corresponding to an edge-weighting of a supertree $T$), we define the distance between  $D^T$ and each $d^t$ by constraining $D^T$ to the rows and columns corresponding to the species that are in $t$ and then computing any of several natural distances between matrices (e.g., the $L_2$ or $L_{\infty}$ metrics). 
The objective is an additive distance matrix that minimizes that total distance.
Once such an additive distance matrix is found, then the tree corresponding to the distance matrix is returned.
%However, finding the optimal supertree under any of these criteria is NP-hard, since the simplest case (when there is just a single source tree) amounts to finding an additive matrix that is closest to the   (see \cite{ailon2011fitting,fakcharoenphol2004tight,agaretal-journal} for an entry into this literature). 

Another approach for distance-based supertree optimization is to compute a single distance matrix $D$ from the  set of distance matrices $\{d^t: t \in \mathcal{T}\}$  and then seek an additive distance matrix $D^*$ that is close (under some metric) to $D$ \cite{lapointe1997average,criscuolo_sdm_2006}.
For example, $D[i,j]$ can be set to the average of the $d^t[i,j]$ for all the source trees $t$ that contain both $i$ and $j$, and the objective could be an additive matrix that minimizes the $L^2$-distance to $D$.
Once an additive matrix is found, the tree (and its set of edge weights) can be computed in polynomial time. 
Thus, this approach can also be used to compute supertrees\footnote{This specific approach is used in   NJst \cite{njst} and ASTRID \cite{astrid}, two distance-based  methods for species tree estimation where the source trees are gene trees (rather than species trees estimated on subsets of the taxon set).
}.

Distance-based tree estimation is a very well-studied problem, and there are many methods that have been developed to construct trees from distances \cite{waterman76}. %Cite books by those french authors?
%Furthermore, under some conditions, it can be shown that the internode distance matrix will be sufficiently close to an additive matrix for the true species tree so that many of the distance-based tree estimation methods will be guaranteed to correctly reconstruct the true species tree from the input distance matrix. 
Fortunately, although finding the nearest additive matrix to a given distance matrix is NP-hard (see \cite{ailon2011fitting,fakcharoenphol2004tight,agaretal-journal} for an entry to this literature), 
 there are many polynomial time methods for computing trees from distance matrices, including neighbor joining \cite{NJ} and FastME \cite{fastme-2015}, and some interesting theory about these approaches \cite{essw1,essw2,DesperGascuel2004,SteelGascuel2006,pardi-bmep,BordewichMihaescu2010,Shigezumi,why-nj-works}.
 %Gascuel and Steel (2006); Fiorini and Joret (2012
However, distance-based tree estimation has another challenge in that for many supertree datasets, there can be pairs of species $i,j$ that are not found together in any source tree, and when this happens $D[i,j]$ will not have any value. 
This is referred to as ``missing data", and 
computing trees from distance matrices with missing data is not well solved.
A few methods are available to estimate trees from distance matrices with missing data \cite{phyd,popescu2012ape,huber-lasso}; however, the accuracy of trees computed from incomplete distance matrices is generally not as good as trees computed from complete matrices \cite{fastrfs}.

In summary, distance-based supertree estimation has been approached in two different (but related) ways: (1) represent each source tree by an additive distance matrix and then seek an additive distance matrix that is close to the set of additive distance matrices computed on the source trees,  or (2) represent the set of source trees by a distance matrix $D_0$ (that may be incomplete) and then seek an additive matrix that is close to $D_0$.
Both approaches are challenged by the fact that desirable definitions of ``close" result in NP-hard optimization problems (and current approaches for finding good solutions to NP-hard optimization problems are not scalable).
The second approach has the additional challenge that the matrices $D_0$ produced by supertree profiles are often incomplete, and current methods for constructing trees from incomplete distance matrices do not have adequate accuracy.
Thus, highly accurate distance-based supertree estimation requires novel techniques in order to scale to most large supertree datasets. 
%Tandy - need citation. Is it even true?

\section{Accuracy and Scalability of Existing Supertree Methods}
%Check quartet-based supertree compared to MRP.
%

Up to 2011,  MRP has been generally found to produce more accurate supertrees than competing methods, 
leading \cite{brinkmeyer2011polynomial} to refer to MRP a ``gold standard" for supertree methods. 
Several supertree methods have been developed since then, some of which have shown to be competitive with MRP in terms of accuracy. 
%Of these more recent methods,  the Bad Clade Deletion \cite{fleischauer2017bad} has good accuracy but requires rooted source trees. 
For example,    ASTRID \cite{astrid},  
ASTRAL-2 \cite{astral-2} (an improved version of ASTRAL \cite{astral}), MRL \cite{mrl}, and FastRFS \cite{fastrfs} all demonstrate high accuracy that is comparable to (or better than) MRP under some conditions.
However, ASTRID has poor accuracy on supertree datasets where there is no source tree containing most of the species, whereas other supertree methods can perform well under those conditions \cite{fastrfs}.

 %MRL, ASTRAL-2, and FastRFS all had comparable accuracy on simulated datasets \cite{fastrfs}, and completed even on the largest supertree datasets examined (one with 2228 species).

As discussed earlier, maximum likelihood and Bayesian supertree estimation methods are also very promising, and 
a maximum likelihood  supertree method  
\cite{Akanni-MLsupertree}  and a Bayesian method  
 \cite{Akanni-Bayesian} have been proposed for supertree estimation under the Steel and Rodrigo model.
 Yet both methods are computationally intensive, and have not (to our knowledge) been used on large datasets.

 %, and can specifically address for and other supertree methods have been developed that are also based on statistical models  \cite{Posada-BayesianSupertree,Ronquist,CottonWilkinson2009}.
 
%\subsection{Computational aspects of existing methods}
Thus, all the methods we have described are computationally intensive on large datasets, and most do not scale to large datasets.
In particular,  nearly all supertree methods explicitly search for good solutions to NP-hard problems using standard techniques for exploring tree space, which are ineffective on large datasets
since the number of trees grows exponentially in the number of leaves.
Some of the distance-based methods are polynomial time (i.e., the ones that first compute a distance matrix summarizing all the source trees, and then run polynomial time distance-based methods such as neighbor joining).
However, the polynomial time distance-based supertree methods have not been shown to provide comparable accuracy to MRP, and furthermore have reduced accuracy when the average distance matrix they compute has missing entries.
Modifications of these distance-based methods so that they also seek good solutions to NP-hard optimization problems might improve accuracy but would also make them computationally intensive.
Hence, scalability to large datasets is not yet achieved using existing supertree methods.

Thus, the best {\em current} supertree methods are either based on encodings using the bipartitions in the source trees (i.e., the MRP matrix) or the quartet trees in the source trees, and
these representations are not efficient.
In particular, the MRP matrix has one row for every species and as many columns as there are edges in the input source trees, and so is larger than the input profile!
For example, suppose we wish to run MRP  on a total of $10,000$ species from  $100$ source trees each with $1,000$ species;  the MRP matrix will have $10,000$ rows and nearly $100,000$ columns. This is a huge matrix, and  maximum parsimony heuristics are not likely to be able to find good local optima on such a dataset.
A similar issue occurs for quartet-based supertree methods, since there are $\Omega(n^4)$ quartets on an $n$-species dataset, and this is quickly a very large number even for modest values for $n$.

%As we have shown, most (but not all) supertree {\em methods} are based on applying standard phylogeny estimation methods to encodings of the input trees, and hence are typically heuristics for  NP-hard optimization problems that use search strategies that are typical of phylogeny estimation methods:  hill-climbing based on standard modifications to tree topologies (and possibly numeric parameters on the tree), randomization to get out of local optima, etc. 
%These search strategies can be highly effective on small datasets, but  do not tend to scale well to very large datasets.
%In particular, some of these methods will fail to find good local optima on datasets containing thousands to tens of thousands of species, and may result in very large running times  even on smaller datasets \cite{fastrfs}. 
%The Bayesian methods use MCMC (Markov Chain Monte Carlo) sampling to explore the space of supertrees, and are also very computationally intensive (arguably more intensive than maximum likelihood methods, since they need to converge to the stationary distribution before their output is deemed reliable).

\section{Improving Scalability of Supertree Methods}

This  section describes some of the major approaches to supertree estimation, focusing on those methods that have been designed for improved scalability to datasets with large numbers of species.

%\subsection{Constrained Optimization}
\subsection{SuperFine: boosting supertree methods}
The SuperFine method \cite{SuperFine} was developed to improve the scalability of supertree methods to large datasets. 
Thus, the input to SuperFine is a set of source trees and also a selected supertree method, such as MRP, MRL, Quartets MaxCut, etc.

\begin{figure}[b]
\includegraphics[scale=.4]{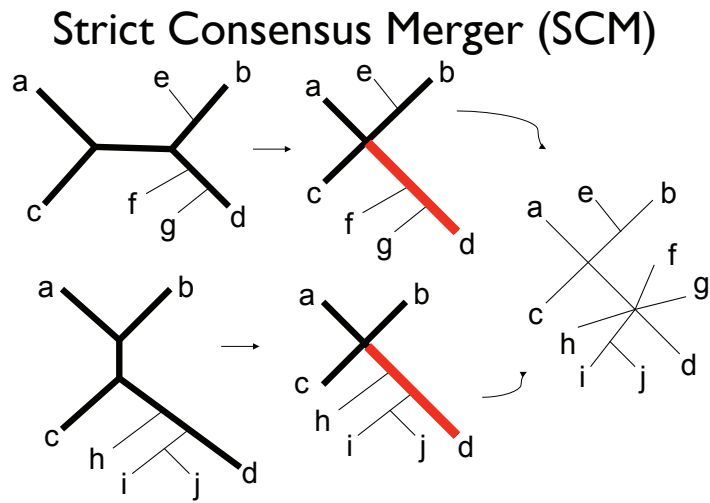}
\caption{(From \cite{warnow-book}) The Strict Consensus Merger (SCM) method, which is the first step in Superfine \cite{SuperFine}. The figure shows how the SCM of two trees is computed. First, the induced trees (shown with thick black lines) on the set of shared taxa in the two trees are computed, and the conflicting edges are collapsed in each tree so that they induce the same subtree  (called the ``backbone tree") on the shared taxa. Then the two trees are merged into a supertree by adding the missing taxa to the backbone tree.  A "collision" occurs when both trees contribute taxa to the same edge in the backbone tree. 
In the event of collision, all subtrees in the two trees contributing to that edge will be attached to the same location in the merged supertree.
Thus, the SCM tree will tend to have lower resolution than the source trees.}
\label{fig:scm}       % Give a unique label
\end{figure}

SuperFine has two steps, where the first step produces a constraint tree, and the second step refines the constraint tree using the base supertree method.
The  constraint tree produced in the first step is an unrooted tree that contains all the species,  called a ``Strict Consensus Merger" tree, that by design only contains bipartitions that are consistent with all the input source trees).
As shown in Figure \ref{fig:scm}, the SCM tree is computed by merging pairs of source trees, until the entire set of source trees is merged into a single tree.
To merge two trees, a ``backbone tree" (the strict consensus of the two trees on their set of shared leaves) is computed; this backbone tree is typically not fully resolved, since any conflict between the two trees will result in the strict consensus being unresolved. 
Then the two trees are used to add the missing taxa into the backbone tree, but this step can also result in lack of resolution if they both contribute taxa to the same edge in the backbone tree. 
Thus, the Strict Consensus Merger (SCM) tree is rarely a binary tree, and can be highly unresolved (i.e., it has nodes of high degree).

The second step of Superfine refines the SCM tree into a binary tree using the base supertree method, applied to encoded versions of the source trees.  
The refinement step has several nice properties.
First,  the polytomies (i.e., nodes in the SCM that have degree four or more) can be refined independently without changing the final supertree; thus, this step can be easily parallelized \cite{Diogo2012,Diogo2017}.
The other nice property is that the refinement of a single node of degree $d$ can be performed by applying the base method to a modified version of the source trees where each source tree is replaced by a new source tree with leaves labelled $1...d$. Therefore, if the maximum degree $d_{max}$ of any node in the SCM is not large (say, at most 100), then the refinement step can be very fast since the base supertree method is only applied to a collection of source trees with a total of $d_{max}$ species.

\begin{figure}
\includegraphics[scale=.65]{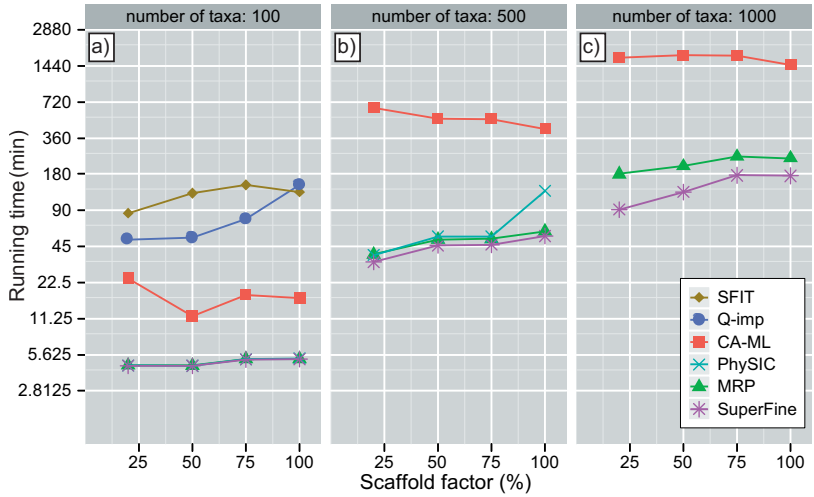}
\caption{(From \cite{SuperFine}) 
We show running time  in minutes for  five supertree methods (log-scale)
and also  concatenation using maximum likelihood (CA-ML) on simulated supertree datasets
with varying scaffold factors (the percentage of species in the largest source tree) for three
numbers of taxa.
%This figure used default positioning, and is too big to use sidecaption. 
}
\label{fig:4}       % Give a unique label
\index{supertree methods! SuperFine}
\end{figure}

\begin{figure}
\includegraphics[scale=.65]{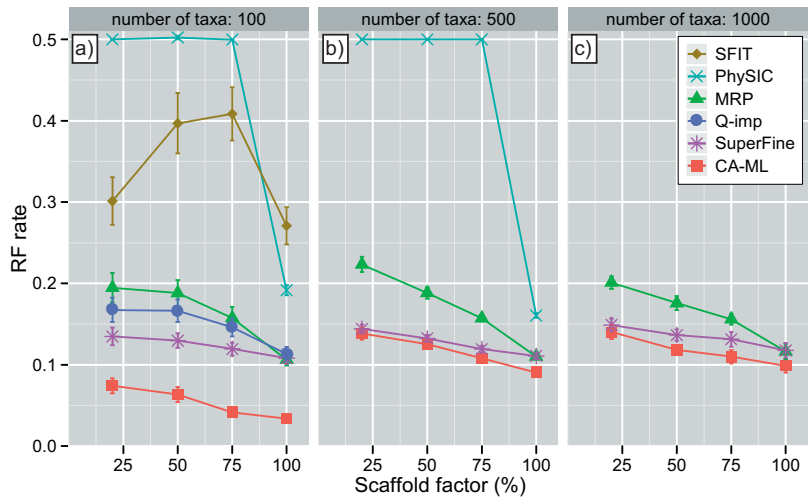}
\caption{(From \cite{SuperFine}) 
We show topological error rates (normalized Robinson-Foulds distances to the true tree) for
five supertree methods and also concatenation using maximum likelihood (CA-ML) on
simulated supertree datasets with varying scaffold factors (roughly the percentage of 
species in the largest source tree) for three numbers of taxa.
%This figure used  default positioning, and is too big to use sidecaption. 
}
\label{fig:5}       % Give a unique label
\index{supertree methods! SuperFine}
\end{figure}

Figures \ref{fig:4} and \ref{fig:5} show a comparison between SuperFine used with MRP and several other supertree methods  and also with ``CA-ML" (the use of maximum likelihood to estimate a tree on the concatenation of the multiple sequence alignments) on a collection of simulated datasets with 100, 500, and 1000 species.
The running time comparison (Figure \ref{fig:4}) shows that SuperFine and MRP have about the same running time on the 100-taxon datasets, but that SuperFine is faster than MRP on the larger datasets. 
Figure \ref{fig:5} shows tree error, using   the normalized Robinson-Foulds \cite{robinson_comparison_1981} between estimated supertrees and the true supertree. 
The x-axis in each subfigure shows the scaffold factor, which is  the percentage of the species that appear in the scaffold source tree (formed by taking a random subset of the taxa). 
Thus, when the scaffold factor is 100\%, at least one of the source trees has all the species.
Note that the scaffold factor impacts the accuracy of the estimated supertree for all methods of computing supertrees, with the most accurate results when the scaffold factor is 100\%.
Most importantly, note that MRP is less accurate than SuperFine using MRP, showing that SuperFine improves accuracy.
Since SuperFine uses MRP to refine the SCM  tree, what this shows is that using the SCM constraint tree is helpful in terms of reducing supertree estimation error.
Furthermore,  as shown in \cite{SuperFine,mrl}, this improvement is not restricted to MRP--SuperFine also improves the speed and accuracy of other supertree methods, including MRL \cite{mrl}  and Quartets MaxCut \cite{snir_quartets_2008}.

However, if the SCM tree is unresolved, then there is no benefit to using SuperFine since the refinement step is identical to just applying the base supertree method to the original set of source trees.
Furthermore, if the SCM tree has very high degree nodes (reflected by having very few internal edges), then the opportunity for improvement is greatly reduced.
On five biological datasets studied in \cite{mrl}, the maximum degree of resolution in the SCM tree was 57\%, where the resolution is the ratio between the number of internal edges in the tree and the   number of internal edges in a fully resolved tree on the same set of leaves. 
In fact, for two of these datasets,  the SCM tree had at most 10\% resolution (one of which only had 1\% resolution). 
SuperFine cannot provide any useful advantage on such datasets.

There are two conditions that cause the SCM to have reduced resolution: one is conflict between the source trees (i.e., two source trees are incompatible), and the other is insufficient overlap between source trees.
Both conditions occur in biological datasets.
In particular, conflict between source trees is inevitable in biological data analysis because source trees are estimated species trees rather than true species trees.
Furthermore,  conflict between pairs of source trees is expected to increase with the number of species and the depth of the evolutionary history, as phylogeny estimation error is known to be impacted by these factors.
In addition, since the SCM is computed by greedily combining source trees until they are all merged into a single tree, the number of source trees should also impact the degree of resolution in the SCM (with less resolution occurring for larger numbers of source trees).
This is consistent with the results shown in  \cite{mrl}, where the SCM trees with the lowest degree of resolution occurred for the datasets with the most source trees, suggesting that this

In summary, although SuperFine can improve supertree methods scalability and accuracy, these improvements only occur when the SCM is highly resolved, and biological supertree  datasets with large numbers of source trees and/or large numbers of species, and especially those spanning deep evolutionary histories, would seem  likely to produce SCM trees that are highly unresolved.
In other words, SuperFine may not enable scalability to  datasets  with many thousands of species, except - perhaps - under limited conditions.

\subsection{Explicitly Constraining the Search Space}
Although SuperFine's technique for constraining the search space doesn't always help, there are other ways that the search space can be constrained. 
Here we describe one such type of approach where  a set $X$ of ``allowed bipartitions" is constructed in some manner, and a supertree $T$ is sought  that optimizes some criterion, subject to the constraint that  $Bip(T) \subseteq X$.
Two supertree methods -- FastRFS \cite{fastrfs} and ASTRAL \cite{astral,astral-2} -- both use this approach.
%FastRFS is a supertree method for the NP-hard Robinson-Foulds Supertree problem that uses this approach, and ASTRAL \cite{astral} is a related method that also uses this approach, and that can also be used for supertree estimation.

Phylogeny estimation using constrained optimization (where the constraints are explicitly provided by a set of allowed bipartitions or clades) was first introduced in \cite{lager} in the context of constructing species trees from a set of gene trees, under a duplication-loss model.
This technique has also been used in other phylogeny estimation methods for different optimization problems \cite{ThanNakhleh2009,YunWarnowNakhleh,BryantSteel-quartets,BayzidPSB,SzollosiBoussau2013,fastrfs,svdquest}.
Each of these methods seeks a tree that optimizes some criterion (typically the distance to the input profile), but constrains the set of feasible trees by an explicitly constructed set $X$ of allowed bipartitions (or, if the tree that is sought must be rooted, then by an explicitly constructed set $X$ of allowed clades). 
Once the set $X$ is defined, the construction of the optimal tree drawing its bipartitions from $X$ can be performed in polynomial time using dynamic programming, although the specific subproblem formulation depends on the optimization problem.

The accuracy of these methods clearly depends on how the set $X$ is defined, since if $X$ is very small then the set of feasible solutions is also very small (or may even be empty), while if $X$ is all possible bipartitions then these methods are guaranteed to find the globally optimal tree.
However, because the running times of these dynamic programming methods grow (polynomially) with $|X|$, it is infeasible to make $|X|$ too large.

The simplest way to define $X$ is to use all the bipartitions of the source trees in $\mathcal{T}$, and this is the technique used in nearly all the methods described above.
However, when a source tree $t$ does not contain all the species, its bipartition set $Bip(t)$ is not directly usable, since these bipartitions are not on $S$ but rather on $\mathcal{L}(t)$, which is a proper subset of $S$.
Hence, this technique is restricted to using just those source trees that contain all the species.

Most of the methods above (with the sole exception of FastRFS) were developed for species tree estimation where the input is a set of gene trees; in that setting, typically some (and perhaps most) of the gene trees will contain all the species.
Furthermore, phylogenomic species tree estimation is increasingly being performed with hundreds to thousands (or tens of thousands) of genes \cite{wickett2014,jarvis-2014b}.
Hence, for the case of phylogenomic species tree estimation, this simple technique will produce a set $X$ that has a large number of bipartitions.
Furthermore, as shown in \cite{YunWarnowNakhleh,astral,svdquest}, highly accurate species trees can be constructed using this technique, making this approach useful in practice for species tree estimation from multi-gene datasets.

However, the supertree setting presents challenges to using this technique to compute the set $X$.
As noted,  when a source tree $t$ does not contain all the species, its bipartition set $Bip(t)$ is not directly usable, and hence this technique is restricted to just those source trees that have all the species. 
But, large supertree datasets may have no such source trees, making this simple technique for computing $X$ completely useless. 
Alternative approaches for constructing the constraint set $X$ of bipartitions when the input profile has no complete source trees have been developed and used in ASTRAL-2 \cite{astral-2} (and further enhanced in ASTRAL-3 \cite{astral3}).
The basic approach in \cite{astral-2,astral3} is to construct a distance matrix relating all the species, and then use that distance matrix to ``complete" all the incomplete gene trees; then, the bipartitions from the completed gene trees can be used for $X$; this is a relatively efficient method on most datasets, but it can be slow on some large datasets with poorly supported gene trees.
%This approach works well in practice in terms of accuracy, and manages to compute all the calculations in $O(n^2)$ time. 
Similarly, \cite{holland-qimp} can be used to complete incomplete source trees, but it runs in $\Omega(n^4)$ time  (where $n$ is the number of species) and so is not scalable to large datasets.
OCTAL \cite{octal} is a linear time algorithm that uses a different approach to complete source trees, but this require that a tree containing all the  species already be available.
Thus, all the approaches for completing incomplete source trees seem to have inherent limitations that reduce scalability. 
In conclusion, defining the constraint set $X$ of allowed bipartitions is challenging for many supertree datasets, and especially so when the number of species is very large. 

However, if an initial set $X$ of allowed bipartitions can be computed using some technique, adding to that set can be beneficial in terms of the resultant supertree or species tree, as shown in \cite{fastrfs,svdquest}.
For example, FastRFS is designed to find optimal solutions to the Robinson-Foulds Supertree problem, within the constraint set defined by the set $X$ of allowed bipartitions.
Since MulRF \cite{mulrf} and PluMiST \cite{plumist-journal} are also designed to find good solutions to the same optimization problems, adding the bipartitions from the trees they find to the initial set $X$ computed by FastRFS ensures that FastRFS will be guaranteed to find solutions that are {\em at least} as good (with respect to the optimization problem) as those found by MulRF and PluMiST. 
Furthermore, adding computed supertrees or species trees, however they are computed, can enlarge the search space and lead to improved topological  accuracy; this is the basis of the ``enhanced" versions of FastRFS and SVDquest, which use bipartitions from trees computed using other methods for species tree estimation, and obtain more accurate species tree as a result.

\section{Using Supertrees in Divide-and-Conquer Strategies}

%Discuss SIESTA, terraces, non-uniqueness.

As we have discussed, large-scale phylogeny estimation presents very substantial challenges, including computational scalability (because the desirable approaches are based on NP-hard optimization problems) and model complexity.
As a result, the estimation of the ``Tree of Life'', a tree that would contain many millions of species, is likely to be extremely difficult to achieve with high accuracy.

One approach that has been proposed \cite{bininda2005supertree,warnow-book} for estimating  large trees is divide-and-conquer, where a dataset is divided into overlapping subsets, trees are estimated on each subset, and then the trees are merged together into a single tree using a supertree method.
There are multiple reasons that such an approach can be beneficial, including reduction in running time, ability to apply different statistical models to different subsets, and ability to use more computationally intensive (and potentially more accurate) methods on each subset.
Indeed, the potential for benefit in large-scale estimation of phylogenetic trees led 
 Bininda-Emonds to assert ``Any attempt to reconstruct large portions of the Tree of Life
will require the use of supertree construction as part of a divide-and-conquer
strategy to phylogenetic reconstruction"  \cite{bininda2005supertree}.

DACTAL (Divide-and-Conquer Trees (almost) without ALignments) \cite{dactal} is an example of how a divide-and-conquer approach can be used for large-scale tree estimation.
The initial motivation for DACTAL was to bypass the challenge of estimating very large alignments, since large datasets are often difficult to align with high accuracy and alignment error can result in tree error.
Thus, DACTAL was developed to enable tree estimation from unaligned sequences, without ever needing to compute a multiple sequence alignment on the entire dataset.
 DACTAL combines divide-and-conquer with iteration, so that each iteration decomposes the input set into subsets using the tree from the prior iteration.

\begin{figure}[h!]
\includegraphics[height=3.5in]{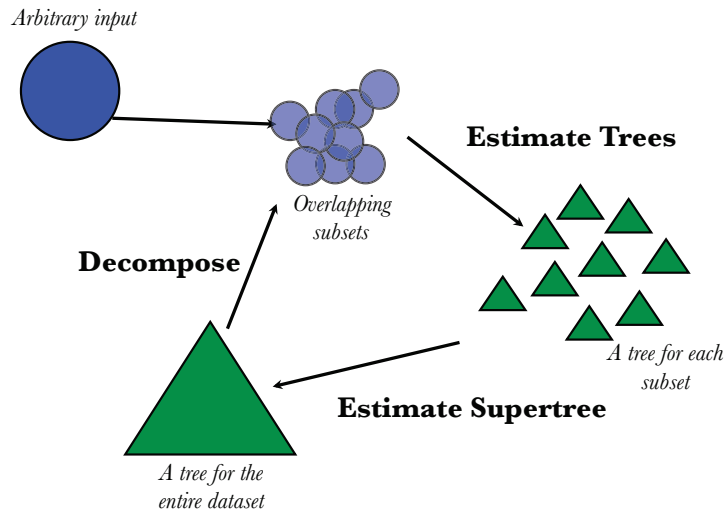}
\caption{(Figure modified from \cite{dactal}) 
The use of divide-and-conquer, as well as iteration, to scale tree estimation methods to large datasets. 
The input is
an arbitrary set of taxa with associated data (e.g., sequences).
In the first step, the dataset is decomposed into overlapping
small subsets of a desired size.
Two ways of
decomposing a sequence dataset into subsets have been developed: one just uses
the sequences, but the other begins by computing a tree (through a
fast but approximate  method).
Then trees are computed
on each subset,  and the subset trees are combined
together using a supertree method.     If desired, the
cycle can then begin again, using the current tree. Each subsequent
iteration begins with the current tree, divides the dataset
into subsets using the tree, computes trees on subsets, and
combines the subset trees using the supertree method.
(This is a modification of the DACTAL \cite{dactal} algorithm, so it
can be used generically.)
\index{algorithms}
\index{algorithms! iteration}
\index{algorithms! divide-and-conquer methods}
\index{DACTAL}
\index{supertree methods}
\index{disk-covering methods (DCMs)}
}
\label{fig:generic-dactal}
\end{figure}

As shown in \cite{dactal}, trees computed using DACTAL were about as accurate as trees computed using SAT\'e \cite{sate2009}, and more accurate than trees computed the best two-phase methods (maximum likelihood on estimated alignments) available at that time.
A generalization of DACTAL (shown in Figure \ref{fig:generic-dactal}) has also been used in phylogenomic species tree estimation  (where species trees are estimated by combining gene trees) and shown to improve the accuracy and speed of MP-EST \cite{mpest}, a coalescent-based species tree method \cite{bayzid-dcm}.

Thus, divide-and-conquer strategies can improve tree estimation accuracy and scalability.
Obviously, the many studies comparing supertree methods \cite{bininda2005supertree,brinkmeyer2011polynomial, SuperFine,fastrfs,mrl,siesta-journal} have shown substantial differences in accuracy between supertree methods (as well as differences in running time), and so the choice of supertree method will impact accuracy and scalability.
Interestingly, how the dataset is divided into subsets will also impact accuracy, with tree-based decompositions generally producing better results than random decompositions \cite{roshan_performance_2004}.
Thus, divide-and-conquer strategies present very substantial opportunities for advances in methods to estimate the Tree of Life, but also present novel challenges that require method development.  

\section{Relationship to Phylogenomic Species Tree Estimation}
The assumption throughout this chapter is that the input profile contains estimated species trees, so that the only cause for the source trees to be different from the true species tree is estimation error.
Yet, supertree methods are also used to combine estimated gene trees into a species tree, which is a different type of analysis and presents different challenges.

Phylogenomic species tree estimation -- i.e., the estimation of species trees using multiple loci taken from the genomes of the different species -- has become a common practice in systematics \cite{jarvis-2014b,wickett2014}. %add other cites
The traditional approach is concatenation, in which the multiple sequence alignments for the different loci are concatenated into one long alignment (called a super-matrix or a super-alignment), and then standard phylogeny estimation methods, such as maximum likelihood, are used to construct a tree on the super-matrix.
Yet, it 
is now well established that gene trees can differ from the species tree due to multiple biological processes, including incomplete lineage sorting, gene duplication and loss, and horizontal gene transfer \cite{maddison1997}, and that concatenation analyses can be statistically inconsistent (and worse, positively misleading) when there is sufficient heterogeneity between true gene trees and species trees \cite{RochSteel}. 
%add cites
Furthermore, gene tree heterogeneity is commonly found in large-scale biological datasets (e.g., \cite{jarvis-2014b,wickett2014}), leading many systematists to seek alternative approaches to concatenation analysis in constructing species trees.

Much of the focus in this area has been on addressing gene tree heterogeneity due to incomplete lineage sorting (ILS), as it is expected to appear in all large phylogenomic datasets to some extent \cite{edwards}.
Furthermore,  gene tree heterogeneity due to ILS is modelled by the multi-species coalescent (MSC), and  there are several methods for estimating species trees that have been proven to be statistically consistent under the MSC. %cite papers!
For example, ``summary methods'', which operate by estimating gene trees and then combining the gene trees into a species tree, form one category of species tree estimation methods that are statistically consistent under the MSC.  
Examples of summary methods  that are statistically consistent under the MSC include ASTRAL (and its improved versions), ASTRID, MP-EST \cite{mpest}, NJst, the population tree in BUCKy \cite{larget-bioinf2010}, and GLASS \cite{glass,iglass}. 
Furthermore, some of these methods  (e.g., ASTRAL) provide very good accuracy in practice and are fast enough to run on large datasets. 

On the face of it, a summary method is just a  supertree method. Yet the two types of methods operate on different assumptions about the cause for heterogeneity in the input profile.
Thus, we distinguish between the two terms to reflect the difference in the assumptions made by the methods: supertree methods assume that differences between source trees and the desired supertree are the result of source tree estimation error, while summary methods assume that biological processes also contribute to these differences.

Two of the methods we discussed in this chapter, ASTRAL and ASTRID, are actually summary methods, and were developed to construct species trees from gene trees that can differ from each other and from the true species tree as the result of ILS.
Furthermore, ASTRAL and ASTRID have been proven statistically consistent under the MSC, while the other supertree methods we discussed either have not been evaluated with respect to statistical consistency or are known to be inconsistent (e.g., MRP).

When gene tree discord is the result of gene duplication and loss, then even gene tree estimation is complicated as each species can have multiple copies of any given gene. 
In that case, usually the set of copies is reduced to a single representative for each species, typically based on inferring ``orthology" -- a difficult and still unsolved problem, as discussed in \cite{kimmen-orthology,Lechner2014, Altenhoff2016}.
When each gene tree has at most one copy of each species, then species tree estimation can be obtained using summary methods that take gene duplication and loss into account; examples of such methods include iGTP \cite{igtp}, DupTree \cite{duptree}, and DynaDup \cite{BayzidPSB}.
Alternatively, the entire set of gene copies can be included in each species, and then a ``gene family tree" can be computed. 
By definition, gene family trees can contain multiple copies of each species, rather than at most one copy of each gene - which is the usual assumption of supertree methods.
Hence, species tree estimation from gene family trees is more complicated, but some methods are able to handle multi-copy genes \cite{mulrf,igtp,Guenomu2017}. 

One of the most interesting methods of this type is \textit{guenomu}, a Bayesian supertree method that addresses heterogeneity between gene trees and the species trees due to multiple biological processes.
As shown in \cite{Posada-BayesianSupertree,Guenomu2017}, \textit{guenomu} has high accuracy compared to several other species tree methods and can run on moderate-sized datasets (a few hundred species); furthermore, \textit{guenomu} can construct species trees from gene family trees (which means that there are multiple copies of the species in each gene tree).  
This performance is very encouraging, especially as there are not many species tree methods that can handle gene family trees.
However, \textit{guenomu} has not been studied on datasets where the source trees are estimated species trees rather than gene trees, and so the accuracy of \textit{guenomu} as a supertree method (in the sense that we use it in this chapter) is not yet known.

\section{Further Reading}
This book chapter focused on a specific challenge in supertree estimation: calculating large supertrees (with thousands of species) from unrooted source trees.
Thus, we explicitly did not discuss supertree construction from rooted source trees, which are also popular.
The  MinCut Supertree \cite{SempleSteel-supertree}, MinFlip Supertree \cite{chen_flipping_2003,chen_improved_2006,chen_minimum-flip_2006}, and PhySIC \cite{ranwez2007physic} are the most well-known supertree methods that work with rooted source trees, but newer methods have  also been developed, some of which may have better accuracy \cite{snir_using_2006,SuperTriplets,Snir-TripletMaxCut2016,fleischauer2017bad}.
For example, Bad Clade Deletion (BCD) supertrees \cite{fleischauer2017bad} has better accuracy than leading supertree methods on rooted source trees \cite{siesta-journal}.

There is also a substantial literature about the theoretical aspects of supertree estimation, some of which is an extension of the results for consensus methods \cite{janowitz2003bioconsensus}. 
When the source trees are unrooted, most optimization problems are NP-hard and the theory (which typically has to do with axiomatic approaches) is generally negative \cite{mcmorris};
however,  when the source trees are rooted some problems become polynomial time and some positive  axiomatic theory can be established 
\cite{bininda-emonds_supertree_2002,thorleyview2003}.

One of the theoretical questions regarding supertree estimation has to do with when a set of source trees is ``decisive", so that there is a unique tree that is compatible with all the source trees.
Determining if a supertree profile is decisive is itself NP-hard, and the probability of being decisive is impacted by missing data \cite{Sanderson2010limits}.
However, there are conditions in which the supertree profile will be decisive and in which the construction of the unique supertree can be performed in polynomial time \cite{warnow2001absolute,dactal}.
Finally, there is a relationship between  decisiveness and  ``phylogenetic terraces", which  occurs when there are multiple trees that have the same optimality score \cite{sanderson2011terraces,sanderson2015impacts,chernomor2016terrace}.

For more of the literature about supertrees, see the book edited by Bininda-Emonds \cite{supertree-book}, covering a wide range of topics about supertree methods and biological dataset analyses using these methods, and  \cite{Clann}, an article about Clann, a software suite of supertree methods, which also provides a good overview of the different algorithmic strategies used in earlier supertree methods.

%\cite{thorleyview2003}.

\begin{acknowledgement}
The author also wishes to thank Pranjal Vachaspati for a careful and thoughtful comments on the manuscript.
%We also thank the anonymous reviewers whose comments were helpful in improving the manuscript.
This paper was supported in part by NSF grant CCF-1535977, but much of the work described in this book chapter was done while the author was part of the CIPRES (www.phylo.org) project, an NSF-funded multi-institutional grant that was initially led by Bernard Moret and then subsequently by the author. 
The first divide-and-conquer methods (DCM-NJ, DACTAL, etc.) were developed with CIPRES support, as were the supertree methods SuperFine and the Strict Consensus Merger that enabled those divide-and-conquer methods to have good performance. 
%The CIPRES project was very much a team, with close collaboration between Bernard Moret, myself, and the many other researchers who participated in the project, most significantly including Mark Holder, Junhyong Kim, Wayne Maddison,  Mark Miller, Satish Rao, and David Swofford.

\end{acknowledgement}

%
%\bibliographystyle{spbasic}

%\bibliographystyle{spmpsci}
%\bibliography{new-warnow2}
%please provide a bibtex file, and use your last name for the name of the bibtex file.  For example, stamatakis.bib or gusfield.bib.
%

%\printindex
\end{document}